\documentclass[showpacs,nofootinbib,reprint,prl,aps,preprintnumbers]{revtex4-1}

\usepackage{amsmath,amssymb}
\usepackage{epsfig}  
\usepackage{graphicx}               % Standard graphics package  
\usepackage{url}
\usepackage{hyperref}
\usepackage{bm}
\usepackage{tikz}
\usepackage[utf8]{inputenc}

\usepackage{pstricks}
\usepackage{color}
\usepackage{slashed}

\newcommand{\nc}{\newcommand}  

\nc{\beq}{\begin{equation}}  
\nc{\eeq}{\end{equation}}  
\nc{\beqa}{\begin{eqnarray}}  
\nc{\eeqa}{\end{eqnarray}}  
\nc{\bit}{\begin{itemize}}  
\nc{\eit}{\end{itemize}}

\begin{document}

\preprint{NSF-KITP-16-042}

\title{Color-octet Companions of a 750 GeV Heavy Pion
}
\author{Yang Bai,$^{a}$ Vernon Barger$^{a,b}$ and Joshua Berger$^{a}$} 
\affiliation{$^{a}$Department of Physics, University of Wisconsin-Madison,
  Madison, Wisconsin 53706, USA  \\
$^{b}$Kavli Institute for Theoretical Physics, University of California, Santa Barbara, CA 93106, USA  
  }
\begin{abstract}
  Color octet bosons are a universal prediction of models in which the 750 GeV diphoton resonance corresponds to a pion of a QCD-like composite sector. We show that the existing searches for dijet and photon plus jet resonances at the LHC constrain single productions of color octet states and can be translated into stringent limits on the 750 GeV diphoton rate. For a minimal $5+\overline{5}$ model, the 750 GeV diphoton signal cross section at the 13 TeV LHC is constrained to be below around 5 fb. Future LHC searches for the photon plus jet resonances can establish evidence of a new color-octet state with 20 fb$^{-1}$ and validate a pion-like explanation for the 750 GeV resonance. 
\end{abstract}
\maketitle

\textbf{\textit{Introduction.}}
The ATLAS and CMS collaborations have recently reported excesses in
their searches for a heavy diphoton resonance at a mass of 750 GeV~\cite{ATLAS-CONF-2016-018,CMS-PAS-EXO-16-018}. If the excesses get confirmed by further data at Run 2 of the LHC, the 750 GeV resonance could be the first particle discovered beyond the Standard Model (SM). The relatively large cross-section of the signal suggests strong dynamics, possibly QCD-like, with its confinement scale at the TeV scale~\cite{Harigaya:2015ezk,Nakai:2015ptz,Franceschini:2015kwy,Low:2015qep,Curtin:2015jcv,Bian:2015kjt,Bai:2015nbs,Craig:2015lra,Franzosi:2016wtl,Harigaya:2016pnu,Draper:2016fsr,Harigaya:2016eol}. The 750 GeV resonance behaves in a similar manner to the neutral pion in the SM QCD sector. It couples to SM gauge bosons via the triangle anomaly such that it can be produced via two gluon partons at the LHC and subsequently decay into two photons. 

In the many proposed composite models~\cite{Harigaya:2015ezk,Nakai:2015ptz,Franceschini:2015kwy,Low:2015qep,Curtin:2015jcv,Bian:2015kjt,Bai:2015nbs,Craig:2015lra,Franzosi:2016wtl,Harigaya:2016pnu,Draper:2016fsr,Harigaya:2016eol}, the properties of the SM gauge singlet corresponding to the 750 GeV particle have been studied extensively, including related $\gamma Z$, $ZZ$ and $WW$ decay channels. These models also predict color-octet and color-triplet bosons because the SM-singlet 750 GeV must couple to gluons. Although the existence of such QCD-charged particles have been mentioned in various studies, their detailed properties and current status based on the existing LHC data have not been examined. In this Letter, we determine the properties of the QCD charged states for a class of models and point out that the existing LHC data can already impose stringent constraints for a large fraction of models. Additional data from the LHC Run 2 will very likely find evidence for the color-octet if the 750 GeV resonance has its origin similar to the neutral pion in the SM.

\begin{figure}[htb!]
  \centering
  \includegraphics[scale=0.25]{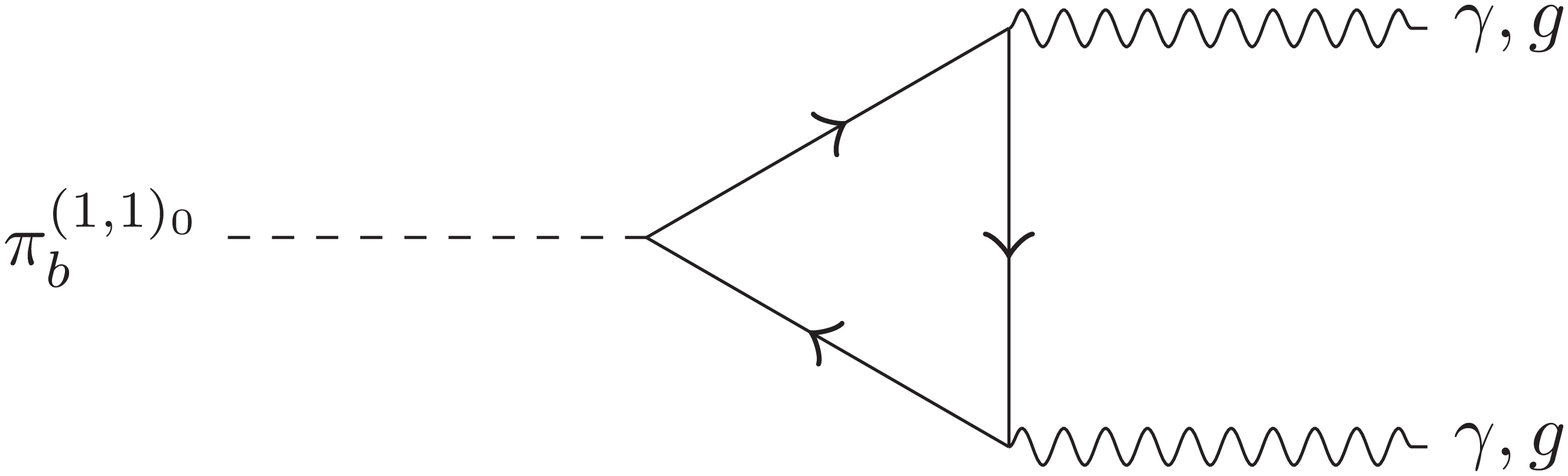} \\
  {\Large $\Downarrow$} \\
  \includegraphics[scale=0.25]{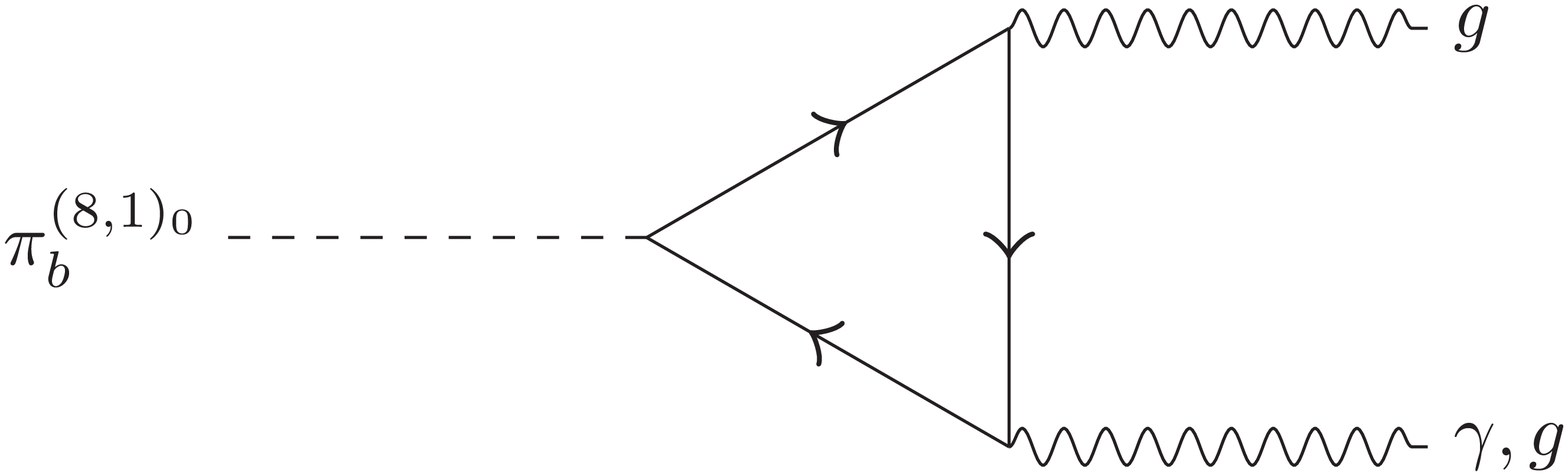}
  \caption{The triangle-anomaly diagrams for the 750 GeV SM-singlet boson (upper) and its predicted color-octet companion (lower) couplings to SM gauge bosons.}
  \label{fig:feyn}
\end{figure}

Unlike the color-triplet heavy pions, whose decays depend on additional higher-dimension operators of undetermined flavor structure, the color-octet heavy pions have leading one pion interactions with SM particles via the triangular interactions with two gluons or one gluon plus one photon as shown in Fig.~\ref{fig:feyn}.  The size of the octet pion couplings are predicted by that of the 750 GeV particle.   The color-octet heavy pions therefore behave as dijet or photon plus jet narrow resonances. In addition to the spin-zero color octet state, the spin-one color-octet vector meson, analogous to the $\rho$ meson, has also been studied. Dijet resonance searches highly constrain this vector meson such that these models are forced into a portion of parameter space in which its dominant decay is to pairs of pions.

 \begin{table*}[tbh!]
  \centering
  \begin{tabular}{c c c}
       \hline\hline
    Model & Big-Quark & Big-pion \\
    \hline
    D $\oplus$ L & $(3,1)_{-1/3} \oplus (1,2)_{1/2}$ & $(8,1)_0
    \oplus (3,2)_{-5/6} \oplus (\bar{3},2)_{5/6} \oplus (1,3)_0 \oplus
    (1,1)_0$ \\
    D $\oplus$ E & $(3,1)_{-1/3} \oplus (1,1)_{1}$ & $(8,1)_0
    \oplus (\bar{3},1)_{4/3} \oplus  (3,1)_{-4/3} \oplus (1,1)_0$ \\
    D $\oplus$ T & $(3,1)_{-1/3} \oplus (1,3)_1$ & $(8,1)_0 \oplus
    (\bar{3},3)_{-2/3} \oplus (3,3)_{2/3} \oplus (1,5)_0 \oplus (1,3)_0 \oplus
    (1,1)_0$ \\
    L $\oplus$ Q & $(1,2)_{1/2} \oplus (\bar{3},2)_{-1/6}$ & $(8,3)_0
    \oplus (8,1)_0 \oplus (3,3)_{2/3} \oplus (\bar{3},3)_{-2/3}
    \oplus$ \\ & & $(3,1)_{2/3} \oplus (\bar{3},1)_{-2/3} \oplus 2\times (1,3)_0 \oplus
    (1,1)_0$ \\
    U $\oplus$ E &  $(3,1)_{2/3} \oplus (1,1)_{-1}$ & $(8,1)_0 \oplus
    (\bar{3},1)_{-5/3} \oplus (3,1)_{5/3} \oplus (1,1)_0$ \\
    U $\oplus$ N & $(3,1)_{2/3} \oplus (1,1)_0$ & $(8,1)_0 \oplus
    (\bar{3},1)_{-2/3} \oplus (3,1)_{2/3} \oplus (1,1)_0$ \\
    E $\oplus$ Q & $(1,1)_{-1} \oplus (3,2)_{1/6}$ & $(8,3)_0 \oplus
    (8,1)_0 \oplus (\bar{3},2)_{5/6} \oplus
    (3,2)_{-5/6}  \oplus (1,3)_0 \oplus (1,1)_0$ \\
    E $\oplus$ S & $(1,1)_{-1} \oplus (6,1)_{2/3}$ & $(27,1)_0 \oplus
    (8,1)_0 \oplus (\bar{6},1)_{1/3} \oplus (6,1)_{-1/3} \oplus
    (1,1)_0$ \\
    S $\oplus$ N & $(6,1)_{2/3} \oplus (1,1)_0$ & $(27,1)_0 \oplus
    (8,1)_0 \oplus (\bar{6},1)_{-2/3} \oplus (6,1)_{2/3} \oplus
    (1,1)_0$ \\
    \hline\hline
  \end{tabular}
        \caption{The representations of big-color quarks and pions under the SM gauge groups $[SU(3)_c, SU(2)_L]_{U(1)_Y}$. The big-quarks are vector-like and have fundamental representations under the big-color gauge group $SU(N_b)$. The SM-singlet big-pion, $(1, 1)_0$, is the 750 GeV diphoton resonance.} 
\label{tab:representations}
\end{table*}

\textbf{\textit{Big-color models for the 750 GeV resonance.}} 
If the strong dynamics giving rise to the 750 GeV resonance is QCD-like, it could arise due to an asymptotically free non-Abelian gauge group, which we denote by $SU(N_b)$.  We call the new color quantum number {\it big-color}. We assume vector-like big-quarks that transform in the fundamental representation of $SU(N_b)$. Some or all of the big-quarks can also be charged under SM gauge interactions. As a phenomenologically broad overview of the possibilities for composite models of the 750 GeV diphoton resonance, we consider the Grand Unified Theory (GUT) inspired models of Ref.~\cite{Redi:2016kip}, and show the SM gauge representations of the big-color quarks and pions for different models in Table~\ref{tab:representations}. 
 
The models in Table~\ref{tab:representations} are selected such that i) they can potentially have SM gauge couplings run up to the GUT scale without hitting a Landau pole (for instance, the $5 + \overline{5}$ or D $\oplus$ L model requires $N_b\leq 10$); ii) the big-quarks 
fall into SM representations that can be embedded in low dimension GUT
multiplets; iii) there is a potential 750 GeV diphoton signal. Within the models considered in Ref.~\cite{Redi:2016kip}, we further select the models that can explain the 750 GeV diphoton resonance using a pseudo-Nambu-Goldstone boson pion as opposed to heavier pseudo-scalar $\eta^\prime$. Models that explain the excess with an $\eta^\prime$ tend to have lighter QCD-charged meson spectra and even tighter constraints than those considered in this work. The SM-singlet meson in each spectrum is fit to the 750 GeV diphoton excess of ATLAS and CMS~\cite{ATLAS-CONF-2016-018,CMS-PAS-EXO-16-018}. There are two hypercharge choices for the color-triplet or color-sextet pions. For convenience, we have chosen one of them in Table~\ref{tab:representations}.

\begin{figure}[htb!]
  \centering
  \includegraphics[scale=0.60]{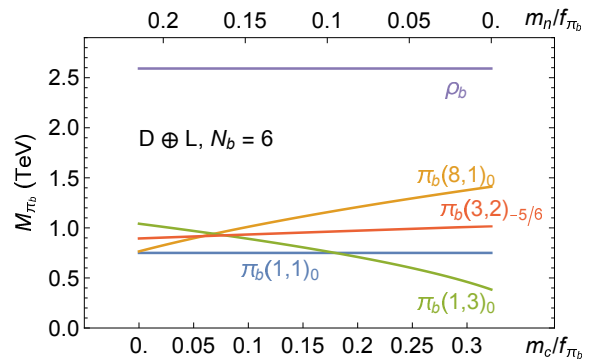}
  \caption{Sample spectrum of light mesons for the D $\oplus$ L model with $N_d = 6$.  We fit the singlet mass to be 750 GeV and fix the pseudo-scalar decay constant such that $\sigma[pp \to \pi_b^{(1,1)_0} \to \gamma\gamma] = 4.6~{\rm fb}$. Here, $m_c$($m_n$) is QCD-charged(singlet) big-quark mass. The upper limit on $m_c/f_{\pi_b}$ arises from requiring $m_n \ge 0$.}
  \label{fig:spectrum}
\end{figure}

All models considered here have a QCD-charged big-quark $\psi_c$ and a QCD-neutral big-quark $\psi_n$.  In the UV theory, we write the big-quark mass terms as
\beq
\mathcal{L} \supset - m_c \overline{\psi}_c \psi_c  -  m_n \overline{\psi}_n \psi_n \,.
\eeq
To determine the constraints on the QCD-charged mesons in the models, we perform a rescaling of the QCD meson spectrum and couplings to determine the structure of our composite sector.  We include both spin-zero big-pions, $\pi_b$s, and spin-one big-rho-mesons, $\rho_b$s, in the spectrum. Their interactions with the SM particles are determined by their decay constants $f_{\pi_b}$ and $f_{\rho_b}$, respectively. The pions obtain their masses from the bare quark masses and the radiative corrections from SM gauge interactions~\cite{Hill:2002ap}. The rho mesons receive small corrections from those two contributions such that all rho mesons could have the same mass at leading order. While the pion spectrum is expected to be accurate at the 10\% level for the parameter space we
consider, the $\rho_b$ meson masses are far more uncertain. In SM QCD, the
fractional mass difference between the lightest and heaviest vector mesons composed of three
light quarks is roughly 30\%.  Since our spectra generally has $m_{c, n} / f_{\pi_b}$
that fall within the SM range, we allow a 30\% uncertainty for the $\rho_b$ masses and take the heavier side in our analysis. After fixing the singlet mass and production cross-section, $4.6\pm 1.2$~fb~\cite{Buttazzo:2015txu}, to fit the 750 GeV diphoton excess, we show a benchmark model spectrum as a function of $m_{c, n}/f_{\pi_b}$ in Fig.~\ref{fig:spectrum}, where we have multiplied our tree-level cross section by a K-factor of 2.5~\cite{Catani:2003zt}. 

The most phenomenologically significant mesons in the short term are those charged under QCD.  Simply by demanding theoretical consistency up to the GUT scale and a fit to the 750 GeV diphoton data, the spectrum of the QCD-charged mesons is highly restricted. For all models considered here that fit the central experimental diphoton cross section of 4.6 fb, the color-octet pion mass is bounded from above by
\beqa
M[\pi_b^{(8, 1)_0}]  \leq 2.2~\mbox{TeV} \,.
\label{eq:octet-mass-bound}
\eeqa
For the minimal $D\oplus L$ model, the upper bound on the color-octet big-pion is 1.7 TeV. If we change the diphoton cross section to be 3.4 fb, the upper bound becomes 1.9 TeV. The
color-triplet pseudo-scalars must be below 1.5 TeV, while the vector $\rho_b$
mesons must be below 4.7 TeV.  Further constraints from
existing searches for these particles apply as we discuss below.

\textbf{\textit{Color-octet (or color-27) big-pions.}} 
\label{sec:octet-pion}
The color-octet big-pion, $\pi_b^{(8,1)_0}$, must be present in all QCD-like models for the 750 GeV resonance. It has the following triangle-anomaly interactions
\beqa
{\mathcal L} &\supset& -\frac{N_b\,\alpha_s }{16\pi \, f_{\pi_b}}\, \frac{A_r}{\sqrt{2\,C_r}}\, d^{abc}\, \pi_b^{8\,a}\,  \epsilon_{\mu\nu\rho\sigma}\, G^{\mu\nu\,b} \, G^{\rho\sigma\, c}   \nonumber \\
&& + \, \frac{N_b\,(3\,Y_c)\,\sqrt{\alpha\,\alpha_s}}{12\pi\,c_W \, f_{\pi_b}}\,\sqrt{2\,C_r}\,\pi_b^{8\,a}\,  \epsilon_{\mu\nu\rho\sigma}\, G^{\mu\nu\,a} \, B^{\rho\sigma} \,.
\eeqa
where $c_W = \cos{\theta_W}$ with $\theta_W$ as the Weinberg angle; $d^{abc}$ is the $SU(3)_c$ symmetric group structure constant; $Y_c$ is the hypercharge of the QCD-charged big-quark; $A_r = 1(7)$ and $C_r = 1/2(5/2)$ for color-triplet(sextet) big-quark. Based on the interactions, the color-octet big-pion can be singly produced via two gluon partons at the LHC. It mainly decays back to two gluons with a smaller but non-negligible branching ratio into $g + \gamma$ given by
\beqa
\frac{\mbox{BR}[\pi_b^{(8, 1)_0} \rightarrow g + \gamma]  }{ \mbox{BR}[\pi_b^{(8, 1)_0} \rightarrow 2 \, g] }  = \frac{8\,\alpha\,(3\, Y_c)^2\,(2 \, C_r)^2}{15\,\alpha_s\,A_r^2}\,.
\eeqa
For the D $\oplus$ L model, the above branching ratio is 4.6\%. The branching ratio into $g + Z$ is similar, but reduced by $s_W^2/c_W^2$; it is $\approx 1.3\%$ for the D $\oplus$ L model. For 4.6 fb of the 750 GeV diphoton signal, the total width of the octet pion is $\approx 135~{\rm MeV}$ for a mass of 1 TeV, so the octet pion is a very narrow resonance. The color-27 pion couples to two gluons via the triangle anomaly; its detailed properties will be presented in Ref.~\cite{BBB-longer}. There is another potential octet big-pion, $(8, 3)_0$, which couples to one gluon plus one weak gauge boson via a triangle anomaly and has interesting pair-production signatures studied in Refs.~\cite{Bai:2010mn,Dobrescu:2011px} but small single production.

\begin{figure}[thb!]
  \centering
  \includegraphics[scale=0.60]{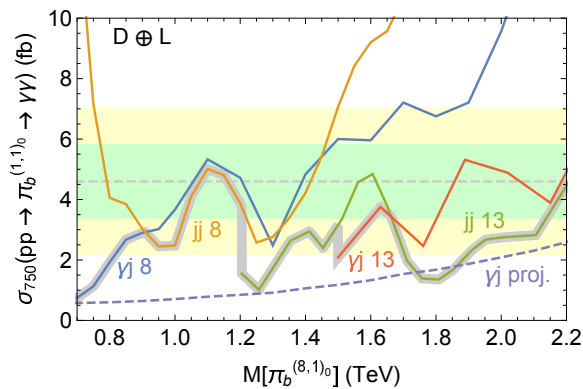}
  \caption{Upper bounds on the 750 GeV diphoton cross section at the 13 TeV LHC for different color-octet pion masses, from various LHC narrow resonance searches including 8 TeV dijet~\cite{Aad:2014aqa}, 8 TeV $\gamma\, j$~\cite{Aad:2013cva}, 13 TeV dijet~\cite{ATLAS:2015nsi} and 13 TeV $\gamma\, j$~\cite{Aad:2015ywd}. Also shown is the our projected limit for $\gamma\, j$ resonance searches with 20 fb$^{-1}$ and 1\% systematic error.}
  \label{fig:sample-limit}
\end{figure}

There are two relevant production channels for the color-octet and color-27 mesons at the LHC: pair production and single resonant production. The pair production channel is determined entirely by QCD couplings to the gluon (up to additional $\rho_b$ mediated productions) and is independent of the decay constant. The mesons decay with nearly 100\% branching fraction to two gluons. A search for pair
production of dijet resonances at the 8 TeV LHC from CMS~\cite{Khachatryan:2014lpa}
currently places a lower bound on the octet mass of 700 GeV and on the
color-27 mass of 1.1 TeV. A correlated channel in which one of the two pair produced mesons
decays to gluon plus photon, rather than two gluons, will likely have
comparable, though somewhat weaker, sensitivity and provides an
important test of such models.

\begin{figure}[thb!]
  \centering
  \includegraphics[scale=0.60]{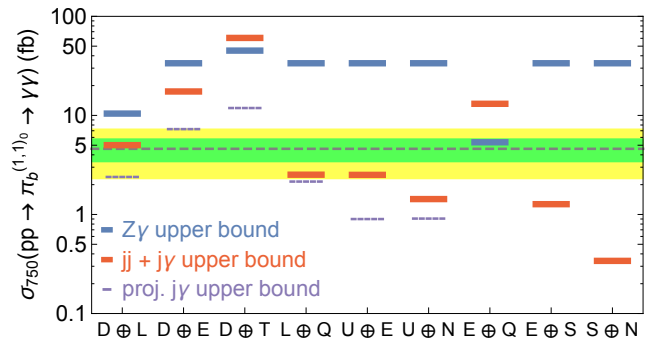}
  \caption{Upper bounds on the 750 GeV diphoton cross section at the 13 TeV LHC from searching for color-octet (and/or color-27) pions in the $jj$ or $j \gamma$ channels. Also shown is the upper bound on the correlated $\pi_b^{(1,1)_0} \to Z\gamma$ channel. All limits assume the weakest constrained color-octet or color-27 mass that is theoretically consistent and not ruled out by pair production searches.}
  \label{fig:current-limit}
\end{figure}

Resonant production depends on the mass and decay constant of the meson. Within a given model and for a given color-octet mass, the ratio $\sigma[p p \to \pi^{(8,1)_0} \to g g] / \sigma[p p \to \pi^{(1,1)_0} \to \gamma \gamma]$ is fixed, so that a bound on dijet resonance production translates directly into a bound on the 750 GeV diphoton cross-section due to the singlet meson. For the D $\oplus$ L model, this ratio is 25/4 for identical color-octet and singlet pion masses. A similar story happens for the color-27 pions~\cite{BBB-longer}. Taking the least constrained color-octet or color-27 meson mass, we place a
conservative upper bound on the 750 GeV diphoton cross-section at the 13 TeV LHC. To derive the projected constraints, we use a Feynrules/MadGraph 5/Pythia 6/PGS~\cite{Alloul:2013bka,Alwall:2011uj,Sjostrand:2000wi,PGS} simulation to estimate the 13 TeV reach with 20 ${\rm fb}^{-1}$ for the octet meson and the correlated singlet cross-section. We have multiplied the color-octet tree-level production cross section by a K-factor of $\approx 3.0$~\cite{Idilbi:2009cc}. In Fig.~\ref{fig:sample-limit} and for the D $\oplus$ L model, we show the upper bounds on the 750 GeV diphoton cross sections for different color-octet masses. One can see that the dijet and $\gamma\, j$ searches are complimentary. The limits are independent of the big-color rank $N_b$. For the theoretically allowed range of octet pion masses in Eq.~(\ref{eq:octet-mass-bound}), the 750 GeV diphoton cross section has an upper bound abound of 5.0 fb, the weakest bound at 1.1 TeV in Fig.~\ref{fig:sample-limit}, which is very close to the central value, 4.6 fb, to fit the diphoton excess~\cite{Buttazzo:2015txu}. It is also interesting to note that there is a small excess at around 1.1 TeV from searches for narrow dijet resonances by ATLAS~\cite{Aad:2014aqa}. In Fig.~\ref{fig:current-limit}, we show the limits for different models by taking the least constrained diphoton cross sections for different color-octet (or color-27) meson masses. 

\begin{figure}[!htb]
  \centering
  \includegraphics[scale=0.6]{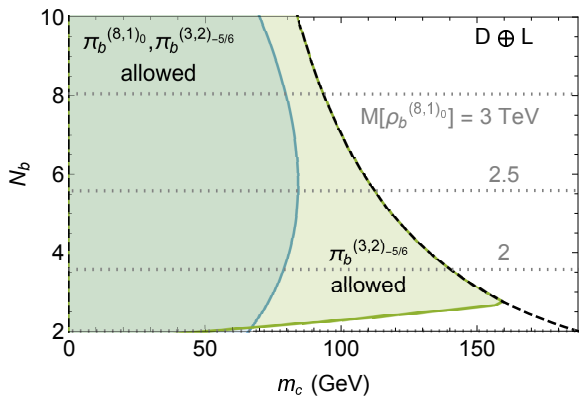}
  \caption{Allowed decay modes for the $\rho_b^{(8,1)_0}$ meson in the D $\oplus$ L  model.  The black dashed line indicates the full limits of the allowed parameter space, such that only decays to $q\bar{q}$ are allowed in the clear region enclosed by the dashed line.}
    \label{fig:phase-space}
\end{figure}

\textbf{\textit{Color-octet $\rho_b$ meson.}} 
\label{sec:rho}
The spin-one color-octet $\rho_b$ meson couples to two QCD-charged big-pions just as the $\rho$ meson couples to pions in the SM. For instance, one has the interaction of $g_{\rho_b \pi_b \pi_b} f^{abc} \rho^a_{b\,\mu} \pi_b^b \partial^\mu \pi_b^c$ for the color-octet pion. The $\rho$--$\pi$--$\pi$ coupling $g_{\rho\pi\pi}$ is estimated using the KSRF relation~\cite{Kawarabayashi:1966kd,Riazuddin:1966sw}. If the two-pion phase space is open, it can decay into two big-pions with the decay width
\beqa
\Gamma\left[\rho_b^{(8,1)_0} \rightarrow \pi_b^{r} \pi_b^{r} \right]
%&& \nonumber \\ &&\hspace{-2cm} 
 = \frac{g_{\rho_b \pi_b \pi_b}^2\,C_r\,M_{\rho_b^{(8,1)_0}} }{96\,\pi} \, \left( 1 -  \frac{4 M^2_{\pi_b^{r} } }{M^2_{\rho_b^{(8,1)_0} }}\right)^{3/2} \,,
\eeqa
with ``$r$" denoting the QCD representation of the big-pion. For $g_{\rho_b \pi_b \pi_b} \approx 6$ and only decay into two octet pions with $C_r = 3$, obtain a decay width of $445$~GeV for $M[\rho_b^{(8,1)_0}]=3$~TeV and $M[\pi_b^{(8,1)_0}]=1$~TeV, which is a very broad vector meson. In Fig.~\ref{fig:phase-space}, we show the allowed region for the vector meson decaying into color-triplet pions with (darker green) or without (lighter green) color-octet pions. 

\begin{figure}[!htb]
  \centering
  \includegraphics[scale=0.6]{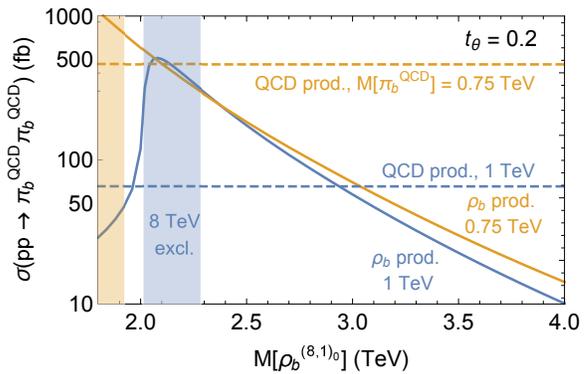}
  \caption{Pair production cross-section of QCD-charged $\pi_b$ at the 13 TeV LHC. The shaded regions are already excluded by searches for pair produced dijet resonances with the blue (orange) shaded region corresponding to a 1 TeV (0.75 TeV) $\pi_b^{\rm QCD}$~\cite{Khachatryan:2014lpa}.}
  \label{fig:vector-cross-section}
\end{figure}

If the two-pion phase space is closed, the $\rho_b^{(8,1)_0}$ meson mainly decays into two SM fermions. The relevant coupling is related to the kinetic mixing between $\rho_b^{(8,1)_0}$ and the QCD gluon and is given by $ i g_s \, t_\theta \, \rho^a_{b\, \mu} \overline{q} t^a \gamma^\mu q$ with $t_\theta \approx 0.2$ as a benchmark point. Ignoring the SM quark mass, the total decay width into all six quarks is  $\Gamma[\rho_b^{(8,1)_0}]= \alpha_s \, t_\theta^2 \,M[\rho_b^{(8,1)_0}] \approx  10$~GeV for  $M[\rho_b^{(8,1)_0}] = 3$~TeV, and hence a narrow resonance. For a $1~{\rm TeV}$ $\rho_b^{(8,1)_0}$ meson, this gives a large dijet production cross-section such that existing dijet resonance searches~\cite{Aad:2014aqa,ATLAS:2015nsi} rule out all models considered other than D $\oplus$ E and D $\oplus$ T. We therefore focus on the portion of parameter space, if it exists, where decays into pseudo-scalar mesons are allowed. In Fig.~\ref{fig:vector-cross-section}, we show the $\rho_b^{(8,1)_0}$ meson-mediated production for QCD-charged pions at the 13 TeV LHC in the Breit-Wigner approximation, which can boost the potential discovery of color-octet pions in the searches for pair produced dijet resonances~\cite{Dobrescu:2007yp,Kilic:2008pm,Bai:2011mr,Simmons:2013zoa}.

\textbf{\textit{Discussion and conclusions.}} 
\label{sec:conclusions}
In all of the models considered here, there are additional QCD-triplet (or sextet) heavy pions. These triplets do not decay without introducing additional non-renormalizable interactions. The lowest dimension-six four-fermion interactions cause leptoquark- or diquark-like decays. The flavor structure of their couplings could have interesting phenomenological consequence in the $B$ or $D$-meson system~\cite{Buttazzo:2016kid}.

We also stress the importance of continuing searches for moderately heavy narrow resonances with smaller cross sections at the LHC. As can be seen from Fig.~\ref{fig:sample-limit}, the search for a $\gamma\,j$ resonance around one TeV is crucial to understand the underlying dynamics of the 750 GeV resonance. One could also look for this color-octet big-pion in the dijet channel using clever ways to overcome trigger and systematics issues~\cite{Dobrescu:2013coa,scouting}. 

In conclusion, color-octet companions of the 750 GeV heavy pion could be discovered by the LHC Run 2 in the dijet or jet plus photon channel, as well as in the four-jet channel from its QCD and vector-meson mediated interactions. The existence of a new and QCD-like strong dynamics can be discovered. 

\vspace{2mm}
This work is supported by the U. S. Department of Energy under the contract DE-FG-02-95ER40896 and in part by the National Science Foundation under Grant No. NSF PHY11-25915. 
\bibliographystyle{JHEP}
\bibliography{OctetRefs}
\end{document}